\documentstyle[12pt]{article} 
  \topmargin - 0.5 cm 
\begin{document}

\centerline{\bf ON TWO APPROACHES TO}

\centerline{\bf FRACTIONAL SUPERSYMMETRIC QUANTUM MECHANICS} 

\bigskip
\bigskip

\centerline{M.~DAOUD\footnote{Current address: 
Laboratoire de Physique de la Mati\`ere Condens\'ee, 
Facult\'e des Sciences, Universit\'e Ibn Zohr, 
BP 28/S, Agadir, Morocco} AND M.~KIBLER}

\medskip
\medskip

\centerline{Institut de Physique Nucl\'eaire de Lyon,}

\centerline{IN2P3-CNRS et 
Universit\'e Claude Bernard,}

\centerline{43 Bd du 11 Novembre 1918, 
F-69622 Villeurbanne Cedex, France}

\begin{abstract}
Two complementary approaches of ${\cal N} = 2$ fractional
supersymmetric quantum mechanics of order $k$ are studied in
this article. The first one, based on a generalized
Weyl-Heisenberg algebra $W_k$ (that comprizes the affine
quantum algebra $U_q(sl_2)$ with $q^k = 1$ as
a special case), apparently contains solely one 
bosonic degree of freedom. The second one uses
generalized bosonic and
$k$-fermionic degrees of freedom. As an illustration, a
particular emphasis is put on the fractional supersymmetric
oscillator of order $k$. 
\end{abstract}

\section{Introduction}
Since more than three decades, supersymmetry is a concept 
largely used in Physics. It was first introduced in elementary 
particle physics on the basis of the unification of internal
and external symmetries.$^{1}$ Curiously enough, 
this kind of unification has never been used directly in Physics
but its corollary according to which the Poincar\'e group has to 
be replaced by an extended Poincar\'e group proved to be fruitful. 
In this respect, the fact that fermions and bosons can be accomodated 
in a given irreducible representation of the $Z_2$-graded Poincar\'e
group is at the origin of the idea of a superparticle.$^{2}$

The experimental evidence for supersymmetry is not yet firmly established. Some 
arguments in favour of supersymmetry come from: (i) condensed matter physics
with
the fractional quantum Hall effect 
and high temperature superconductivity;$^{3}$
(ii) nuclear physics where supersymmetry could
connect the complex structure of odd-odd nuclei to 
much simpler even-even and odd-$A$ systems;$^{4}$
and (iii) high energy physics 
 (especially in the search of supersymmetric particles and the lighter
 Higgs boson)
 where the recently observed signal at 115 GeV$/c^2$ for
a neutral Higgs boson is compatible with the hypotheses of 
supersymmetry.$^{5}$

In spite of the absence of a decisive evidence 
for supersymmetry, supersymmetric quantum
mechanics (SSQM), 
 a supersymmetric quantum field theory in $D = 1+0$ dimension [6], 
has received a great deal of attention in the last
 twenty years and is still in a state of development. 
In recent years, the investigation of quantum groups, 
with deformation parameters taken as roots of unity, 
has been a catalyst for the study of fractional 
supersymmetric quantum mechanics (FSSQM) which is an extension 
of ordinary SSQM. This extension
takes its motivation in the so-called intermediate or exotic statistics like:
(i) anyonic statistics in $D = 1 + 2$ dimensions connected to 
braid groups,$^{3,7-9}$
(ii) para-bosonic and para-fermionic 
 statistics in $D = 1 + 3$ dimensions connected to 
 permutation groups,$^{10-13}$ 
             and (iii) $q$-deformed statistics (see, for instance,$^{14,15}$ 
             arising from $q$-deformed oscillator algebras.$^{16-20}$ Along
this vein,
             intermediate statistics constitute a
  useful tool for the study of physical phenomena in condensed matter physics
(e.g.,
  fractional quantum Hall effect and supraconductivity at high critical
temperature).  

Supersymmetric quantum mechanics needs two degrees of freedom: one bosonic
degree (described by a complex variable) and one fermionic degree (described by
a
Grassmann variable). From a mathematical point of view, we then have a
$Z_2$-grading
of the Hilbert space of physical states (involving 
bosonic and fermionic states). Fractional 
supersymmetric quantum mechanics of order $k$
is an extension of ordinary SSQM for which the
$Z_2$-grading is replaced by a $Z_k$-grading with 
$k \in {\bf N} \setminus \{ 0 , 1 \}$.
The $Z_k$-grading corresponds to a bosonic degree of freedom (described again
by a
complex variable) and a para-fermionic or 
$k$-fermionic degree of freedom (described by a generalized
Grassmann variable of order $k$). In other words, to pass from 
ordinary   supersymmetry or SSQM to 
fractional supersymmetry or FSSQM of order
$k$, we retain the bosonic variable and replace the fermionic variable by a
para-fermionic or $k$-fermionic variable (which can 
be represented by a $k \times k$ matrix). 

A possible approach to FSSQM of order $k$ thus 
amounts to replace fermions by para-fermions 
of order $k-1$. This
yields para-supersymmetric quantum mechanics as first developed, 
with one boson and one para-fermion of order 2, by Rubakov and 
Spiridonov$^{21}$ and extended by various authors.$^{22-27}$ An 
alternative approach to 
FSSQM of order $k$ consists in replacing fermions by $k$-fermions 
which are objects interpolating
between bosons (for $k \to \infty$) and fermions (for $k=2$) and which satisfy
a
generalized Pauli exclusion principle 
according to which one cannot put more than $k-1$ particles 
on a given quantum state.$^{28}$ The $k$-fermions 
proved to be useful in describing 
Bose-Einstein condensation in low dimensions$^{14}$
(see also Ref.~[15]). They take their origin 
in a pair of $q$- and ${\bar q}$-oscillator 
algebras (or $q$- and 
                      ${\bar q}$-uon algebras) with
$$
q = \frac{1}{{\bar q}} := 
    \exp \left( \frac{2 \pi {\rm i}}{k} \right),
\eqno (1)
$$
where $k \in {\bf N} \setminus \{ 0 , 1 \}$. Along this line, a fractional
supersymmetric oscillator was derived in terms of boson and $k$-fermion
operators 
in Ref.~[29]. 

Fractional supersymmetric quantum mechanics was also developed without an
explicit
introduction of $k$-fermionic degrees of freedom.$^{30,31}$ In
this respect, FSSQM of order $k=3$ 
was worked out by Quesne and Vansteenkiste$^{31}$ owing to  
the introduction of the the $C_{\lambda}$-extended
oscillator algebra. Their work is an extension 
of the construction by Plyushchay$^{30}$
of SSQM, viz., FSSQM of order $k=2$, with one bosonic degree of freedom only.

The connection between FSSQM (and thus SSQM) and quantum groups 
has been worked out by several authors$^{32-40}$
mainly with applications to exotic statistics in mind. In
 particular, LeClair and Vafa$^{32}$ studied 
the isomorphism between the affine quantum algebra 
$U_q(sl_2)$ and ${\cal N} = 2$ FSSQM in $D = 1+1$ dimensions when $q^2$
goes to a root of unity (${\cal N}$ is the number of 
supercharges); in the special case where $q^2 \to -1$,
they recovered ordinary SSQM.

It is the aim of this paper to approach 
${\cal N}=2$ FSSQM of order $k$ from 
different routes: (i) first,
from a generalized Weyl-Heisenberg algebra $W_k$ (defined in Sec.~2)
and 
(ii) second, in terms of generalized bosonic and $k$-fermionic operators
(Secs.~4 and 5). In
Sec.~3,  a fractional supersymmetric Hamiltonian is derived 
from the generators of $W_k$ and specialized to 
the case of a fractional supersymmetric oscillator. In Sec.~5, 
this fractional supersymmetric oscillator is further investigated on the basis
of a $Q$-uon approach to the algebra $W_k$, with $Q$ going to a 
$k$-th root of unity. Finally, differential realisations, involving
bosonic and generalized Grassmannian variables, of FSSQM 
are given in Sec.~6 for some particular cases of $W_k$. Three appendices 
complete this paper: The quantum algebra $U_q(sl_2)$ with $q^k = 1$ is
connected to $W_k$ in Appendix A, a boson $+$ $k$-fermion decomposition 
of a $Q$-uon for $Q \to q$ 
is derived in Appendix B and the action of deformed-boson and $k$-fermion
operators on the $Z_k$-graded Hilbert-Fock space is given in Appendix C.

In the present paper, we use the notation $[A , B]_Q := AB - QBA$
for any complex number $Q$ and any pair of operators $A$ and $B$. 
As particular cases, we have $[A , B]$ or 
$[ A , B ]_- := [A , B]_1$ and 
$[ A , B ]_+ := [A , B]_{-1}$ for the commutator and the anti-commutator, 
respectively, of $A$ and $B$. 
 
\section{A generalized Weyl-Heisenberg algebra $W_k$}
\subsection{The algebra}
For fixed $k$, with $k \in {\bf N} \setminus \{ 0 , 1 \}$, we define a  
generalized Weyl-Heisenberg algebra, denoted as $W_k$, as an algebra 
spanned by four linear operators $X_-$ (annihilation operator), 
$X_+$ (creation operator), $N$ (number operator) and 
$K$ (grading operator) acting on some Hilbert space 
and satisfying the following relations:
$$
 [X_- , X_+] = \sum_{s=0}^{k-1} f_s(N) \> \Pi_s, 
 \eqno (2{\rm a})
$$
$$
 [N , X_-]   = - X_-,  \quad
 [N , X_+]   = + X_+,
 \eqno (2{\rm b})
$$
$$
 [K , X_+]_q = [K , X_-]_{\bar q} = 0,
 \eqno (2{\rm c})
$$
$$
 [K , N] = 0, 
 \eqno (2{\rm d})
$$
$$
 K^k = 1,
 \eqno (2{\rm e})
$$
where $q$ is the $k$-th root of unity given by (1).
In Eq.~(2a), 
the $f_s$ are reasonable functions (see below) and the
operators $\Pi_{s}$ are polynomials in $K$ defined by  
$$
\Pi_{s} := \frac{1}{k}  \>  \sum_{t=0}^{k-1}  \>  q^{-st} \> K^t
\eqno (3)
$$
for $s = 0, 1, \cdots, k-1$. Furthermore, we suppose that the operator 
$K$ is unitary ($K^{\dagger} = K^{-1}$), the operator $N$ is self-adjoint
($N^{\dagger} = N$), and the operators $X_-$ and $X_+$ are connected via
Hermitean conjugation ($X_-^{\dagger} = X_+$).  The functions 
$f_s : N \mapsto f_s(N)$ must satisfy the
constrain relation
$$
f_s(N)^{\dagger} = f_s(N)
$$
(with $s = 0, 1, \cdots, k-1$) in order that $X_+ = X_-^{\dagger}$ be
verified.  

\subsection{Projection operators}
It is clear that we have the resolution of the identity operator
$$
\sum_{s=0}^{k-1} \Pi_s = 1
$$
and the idempotency relation
$$
\Pi_s \Pi_t = \delta (s,t) \Pi_s
$$
where $\delta$ is the Kronecker symbol. Consequently, 
the $k$ self-adjoint operators $\Pi_s$ are 
projection operators for the cyclic group 
$Z_k = \{ 1, K, \cdots, K^{k-1} \}$ of order $k$ spanned by the generator $K$.
In 
addition, these projection operators satisfy 
$$
\Pi_s X_+ = X_+ \Pi_{s - 1}  \Leftrightarrow X_- \Pi_s = \Pi_{s - 1} X_-
\eqno (4)
$$
with the convention $\Pi_{-1} \equiv \Pi_{k-1}$ and 
                    $\Pi_k    \equiv \Pi_0$ 
(more generally, $\Pi_{s+kn}  \equiv \Pi_{s}$ for $n \in {\bf Z}$). Note that
Eq.~(3) can be reversed in the form
$$
K^t = \sum_{s=0}^{k-1} q^{ts} \> \Pi_s
$$
with $t = 0, 1, \cdots, k-1$.

\subsection{Representation}
We now consider an Hilbertean representation of the algebra $W_k$. Let
${\cal F}$ be the Hilbert-Fock space 
on which the generators of $W_k$ act. 
Since $K$ obeys the cyclicity condition $K^k = 1$, the 
operator $K$ admits the set $\{ 1, q, \cdots, q^{k-1} \}$ of eigenvalues. It
thus makes it possible to graduate, via a $Z_k$-grading, the 
representation space ${\cal F}$ of the algebra $W_k$ as
$$
 {\cal F} = \bigoplus_{s=0}^{k-1} {\cal F}_s 
 \eqno (5{\rm a})
$$  
where
$$
{\cal F}_s := \{ | k n + s \rangle : n \in {\bf N} \}
 \eqno (5{\rm b})
$$
with
$$
K | k n + s \rangle = q^{s} | k n + s \rangle.
$$
Therefore, to each eigenvalue $q^{s}$ (with $s = 0, 1, \cdots, k-1$)
we associate a subspace ${\cal F}_s$ of ${\cal F}$. It is evident that
$$
\Pi_s | k n + t \rangle = \delta (s,t) \> | k n + s \rangle
$$
and, thus, the application $\Pi_s : {\cal F} \to {\cal F}_s$ yields a
projection of ${\cal F}$ onto its subspace ${\cal F}_s$.

The action of $X_{\pm}$ and $N$ on ${\cal F}$ can be taken to be 
$$
N | k n + s \rangle = 
n | k n + s \rangle
$$
and
$$
X_- | kn + s \rangle = \sqrt {F (n)} \>
                       | k(n-1) + s-1 \rangle, \quad s \not=0,
\eqno (6{\rm a})
$$
$$
X_- | kn     \rangle = \sqrt {F (n)} \>
                       | k(n-1) + k-1 \rangle, \quad s     =0,
\eqno (6{\rm b})
$$
$$
X_+ | kn + s \rangle = \sqrt {F (n + 1)} \>
                       | k(n+1) + s+1 \rangle, \quad s \not=k-1,
\eqno (6{\rm c})
$$
$$
X_+ | kn +k-1\rangle = \sqrt {F (n + 1)} \>
                       | k(n+1)       \rangle, \quad s     =k-1.
\eqno (6{\rm d})
$$
The function $F$ is a structure function that
fulfills the initial condition $F(0) = 0$ (see Refs.~[41,42]). Furthermore, it
satisfies
$$
X_- X_+ = F(N + 1), \quad X_+ X_- = F(N)
$$
and
$$
F(N + 1) - F(N) = \sum_{s=0}^{k-1} f_s(N) \> \Pi_s
$$
which admits the classical solution $F(N) = N$ for $f_s=1$ ($s= 0, 1, \cdots,
k-1$).

\subsection{Particular cases}
The algebra $W_k$ covers a great number of situations
encountered in the 
literature.$^{29-31,43,44}$ These situations differ 
by the form given to the
right-hand side of (2a) and can be classified as follows.

(i) As a particular case, the algebra $W_2$ for $k=2$ with 
$$
[ X_- , X_+] = 1 + c \> K, \quad 
 [N , X_{\pm}]   = {\pm} X_{\pm},
$$
$$
 [ K , X_{\pm} ]_+ = 0, \quad 
 [K , N] = 0, \quad K^2 = 1,
$$
where $c$ is a real constant ($f_0 = 1 + c$, 
                              $f_1 = 1 - c$),
corresponds to the Calogero-Vasiliev$^{43}$ algebra
                considered by Gazeau$^{44}$
for describing a system of two anyons, 
with  an  Sl($2 , {\bf R}$) dynamical symmetry,  
subjected to an intense magnetic field
and by Plyushchay$^{30}$
for constructing SSQM without fermions. Of 
course, for $k=2$ and $c=0$ we recover 
the algebra describing the ordinary or
$Z_2$-graded supersymmetric oscillator.

If we define
$$
c_s = \frac{1}{k} \sum_{t = 0}^{k-1} q^{- ts} \> f_t(N),
\eqno (7)
$$
with the functions $f_t$ chosen in such a way that $c_s$ is independent 
of $N$ (for $s = 0, 1, \cdots, k-1$), the algebra $W_k$ defined
by
$$
[ X_- , X_+] = \sum_{s = 0}^{k-1} c_s \> K^s,
\eqno (8)
$$
together with Eqs.~(2b)-(2e), corresponds to the $C_{\lambda}$-extented 
harmonic oscillator algebra introduced by 
Quesne and Vansteenkiste$^{31}$ for formulating 
FSSQM of order 3. The latter algebra was explored
in great detail in the case $k=3$.$^{31}$

(ii) Going back to the general case where $k \in {\bf N} \setminus \{ 0 , 1
\}$,
if we assume in Eq.~(2a) that $f_s = G$ is independent of $s$ 
with $G(N)^{\dagger} = G(N)$, we get
$$
[ X_- , X_+] = G(N).
\eqno (9)
$$
We refer 
the algebra $W_k$ defined by Eq.~(9) together with Eqs.~(2b)-(2e)
to as a nonlinear Weyl-Heisenberg algebra (see also Ref.~[13]). The 
latter algebra was considered by the authors as a generalization 
of the $Z_k$-graded Weyl-Heisenberg algebra describing a generalized fractional
supersymmetric oscillator.$^{29}$

(iii) As a particular case, for $G = 1$ we have
$$
[ X_- , X_+] = 1
\eqno (10)
$$
and here we can take
$$
N := X_+ X_-.
\eqno (11)
$$
The algebra $W_k$ defined by Eqs.~(10) and (11) together with Eqs.~(2b)-(2e)
corresponds to the $Z_k$-graded 
    Weyl-Heisenberg algebra connected to the fractional
    supersymmetric oscillator studied in Ref.~[29]. 

(iv) Finally, it is to be noted that the affine quantum algebra $U_q(sl_2)$
with $q^k = 1$ can be considered as a special case of 
the generalized Weyl-Heisenberg algebra $W_k$ (see Appendix A).
This result is valid for all the representations (studied in Ref.~[45]) 
of the algebra $U_q(sl_2)$.

\section{A general supersymmetric Hamiltonian}
\subsection{Supercharges}
We are now in a position to introduce
supercharges which are basic operators for
the formulation of FSSQM. We define the supercharge
operators $Q_-$ and $Q_+$ by 
$$
Q_- := X_- (1 - \Pi_{1}), 
\eqno (12{\rm a})
$$
$$
Q_+ := X_+ (1 - \Pi_{0}),
\eqno (12{\rm b})
$$
or alternatively 
$$
Q_- := X_- ( \Pi_{2} + \cdots + \Pi_{k-2} + \Pi_{k-1} + \Pi_{0} ), 
\eqno (13{\rm a})
$$
$$
Q_+ := X_+ ( \Pi_{1} + \Pi_{2} + \cdots + \Pi_{k-2} + \Pi_{k-1} ). 
\eqno (13{\rm b})
$$
Indeed, we have here one of $k$, with 
$k \in {\bf N} \setminus \{ 0 , 1 \}$, 
possible equivalent definitions of the
supercharges $Q_-$ and $Q_+$ corresponding to the $k$ circular permutations of
the indices $0, 1, \cdots, k-1$. Obviously, we have the Hermitean conjugation
relation
$$
Q_-^{\dagger} = Q_+.
$$
Thus, our approach corresponds to a ${\cal N} = 2$ formulation of 
FSSQM of order $k$ 
($\frac{\cal N}{2}$ is the number of independent
supercharges). By making use of the commutation relations between the
projection
operators $\Pi_s$ and the shift operators $X_-$ and $X_+$ [see Eqs.~(4)], 
we easily get
$$
Q_-^m = X_-^m ( \Pi_{0}   + 
                \Pi_{m+1} + 
                \Pi_{m+2} + 
                \cdots    + 
                \Pi_{k-1} )
\eqno (14{\rm a})
$$
$$
Q_+^m = X_+^m ( \Pi_{1}         + 
                \Pi_{2}         + 
                \cdots          + 
                \Pi_{k - m - 1} + 
                \Pi_{k - m} )
\eqno (14{\rm b})                
$$
for $m = 0, 1, \cdots, k-1$. By combining Eqs.~(12) or (13) and (14), we obtain
$$
Q_-^k = Q_+^k = 0. 
$$
Hence, the supercharge operators $Q_-$ and $Q_+$ are nilpotent operators of
order
$k$. 

We continue with some relations at the basis of the derivation of a
supersymmetric Hamiltonian. The central relations are 
$$
Q_+ Q_-^m = X_+ X_-^m ( 1 - \Pi_m ) ( \Pi_{0} + \Pi_{m+1} + \cdots + \Pi_{k-1}
) 
\eqno (15{\rm a})
$$
$$
Q_-^m Q_+ = X_-^m X_+ ( 1 - \Pi_0 ) ( \Pi_{m} + \Pi_{m+1} + \cdots + \Pi_{k-1}
) 
\eqno (15{\rm b})
$$
with $m = 0, 1, \cdots, k-1$. From Eqs.~(15), we can derive the following
identities giving $Q_-^m Q_+ Q_-^{\ell}$ with $m + \ell = k - 1$.

(i) We have
$$
Q_+ Q_-^{k-1} = X_+ X_-^{k-1} \Pi_{0}
\eqno (16{\rm a})
$$
$$
Q_-^{k-1} Q_+ = X_-^{k-1} X_+ \Pi_{k-1}
\eqno (16{\rm b})
$$
in the limiting cases corresponding to $(m = 0,   \ell = k - 1)$ and 
                                       $(m = k-1, \ell = 0    )$.

(ii) Furthermore, we have 
$$
Q_-^m Q_+ Q_-^{\ell} = X_-^m X_+ X_-^{\ell} ( \Pi_{0} + \Pi_{k-1} )
\eqno (16{\rm c})
$$
with the conditions $(m \not= 0,   \ell \not= k - 1)$ and 
                    $(m \not= k-1, \ell \not= 0    )$.   

\subsection{The Hamiltonian}                                       
Following Rubakov and Spiridonov,$^{21}$ we consider 
the multilinear relation
$$
Q_- ^{k-1} Q_+  +  Q_- ^{k-2} Q_+ Q_- 
                                      +   \cdots 
                                      +   Q_+ Q_- ^{k-1}
                                      =       Q_- ^{k-2} H,
$$
where $H$ is an operator that depends on the algebra $W_k$. The 
operator $H$ defines the Hamiltonian for
a supersymmetric system  
associated to $W_k$. This dynamical system, that
we shall refer to a fractional or $Z_k$-graded 
supersymmetric system, depends on the functions $f_s$ 
occurring in the definition (1) of $W_k$. By repeated use
of Eqs.~(1) and (16), we find that the most general
expression of $H$ is 
   $$
   H = (k-1) X_+ X_-                                                       
   - \sum_{s = k-2}^{k  } 
     \sum_{t =   2}^{s-1} (t - 1) \> f_{t}(N - s + t )         \Pi_{s}
   $$ 
   $$   
   - \sum_{s =   1}^{k-1} 
     \sum_{t =   s}^{k-1} (t - k) \> f_{t}(N - s + t )         \Pi_{s}
     \eqno (17)
   $$ 
in terms of the product $X_+ X_-$, the operators $\Pi_s$ and the functions
$f_s$. In 
the general case, we can check that 
$$
H^{\dagger} = H
\eqno (18)
$$  
and
$$
[H , Q_-] = [H , Q_+] = 0.
\eqno (19)
$$
Equations (18) and (19) show 
that the two supercharge operators $Q_-$ and $Q_+$ are two (non independent)
constants 
of the motion for the
Hamiltonian system described by the self-adjoint operator $H$. From 
Eqs.~(17-19), it can be seen that 
the Hamiltonian $H$ is a linear combination of the
projection operators $\Pi_{s}$ with coefficients corresponding to isospectral 
Hamiltonians (or supersymmetric partners) associated to the various subspaces 
${\cal F}_s$ with $s=0, 1, \cdots, k-1$. 
 
\subsection{Particular cases}
The general expression (17) for the Hamiltonian $H$ can be particularized to
some 
interesting cases. These cases correspond to the above-mentioned forms of the
generalized Weyl-Heisenberg algebra $W_k$.

(i) In the particular case $k=2$, by taking $f_0 = 1 + c$ and $f_1 = 1 - c$, 
where $c$ is a real constant, the Hamiltonian (17) gives back the one derived
by
Plyushchay$^{30}$ for SSQM.   

More generaly, by restricting the functions $f_t$ in Eq.~(17) 
to constants (independent of $N$) defined by
$$
f_t = \sum_{s=0}^{k-1} q^{ts} \> c_s
$$
in terms of the constants 
$c_s$ (cf.~Eq.~(7)), the so-obtained Hamiltonian $H$
corresponds to the $C_{\lambda}$-oscillator 
fully investigated for $k=3$ in Ref.~[31].

(ii) In the case $f_s = G$ (independent of $s = 0, 1, \cdots, k-1$), i.e., for
a generalized Weyl-Heisenberg algebra $W_k$ defined by (2b)-(2e) and (9), the
Hamiltonian $H$ can be written as 
$$
H   = (k-1) X_+ X_-                                                             
    - \sum_{s=2}^{k  -1}
        \sum_{t=1}^{s  -1}         G(N-t) (1 - \Pi_1 - \Pi_2 - \cdots -
\Pi_{s})
$$
$$    
    + \sum_{s=1}^{k  -1}
        \sum_{t=0}^{k-s-1} (k-s-t) G(N+t) \> \Pi_{s}. 
        \eqno (20)
$$
The latter expression was derived in Ref.~[29].

(iii) If $G=1$, i.e., for a Weyl-Heisenberg algebra defined by (2b)-(2e) and
(10), Eq.~(20) leads to the Hamiltonian 
$$
H = (k-1) X_+ X_-  
  + (k-1) \sum_{s=0}^{k-1} ( s + 1 - \frac{1}{2} k ) \Pi_{k-s}
\eqno (21)
$$
for a fractional supersymmetric oscillator. The energy spectrum of $H$ is
made of equally spaced levels with a ground state (singlet), a first
excited state (doublet), a second excited state (triplet), $\cdots$, 
a ($k-2$)-th excited state (($k-1$)-plet) followed by an infinite sequence of 
further excited states (all $k$-plets). 

(iv) In the case where the algebra 
$W_k$ is restricted to $U_q(sl_2)$, see Appendix A, the corresponding
Hamiltonian $H$ is given by Eq.~(17) where the $f_{t}$ are given in Appendix A. 
This yields 
$$
 H = (k-1) J_+ J_-                                                       
   + \frac{1}{\sin {\frac{2 \pi  }{k}}} \sum_{s = k-2}^{k  } 
       \sum_{t =   2}^{s-1} (t - 1) \> \sin {\frac{4 \pi t}{k}}  \Pi_{s}   
$$
$$   
   + \frac{1}{\sin {\frac{2 \pi  }{k}}} \sum_{s =   1}^{k-1} 
       \sum_{t =   s}^{k-1} (t - k) \> \sin {\frac{4 \pi t}{k}}  \Pi_{s}.
       \eqno (22)
$$ 
Alternatively, Eq.~(22) can be rewritten 
in the form (20) where $X_{\pm} \equiv J_{\pm}$ and $N \equiv  J_3$ 
and where the function $G$ is defined by
$$
G(X) := - [2X]_q,
$$
where the symbol $[ \ ]_q$ is defined by 
$$
[2X]_q := \frac{ q^{2X}  -  q^{-2X} }{ q - q^{-1} }
$$
with $X$ an arbitrary operator or number. The 
quadratic term $J_+J_-$ can be expressed in term of the Casimir operator $J^2$
of $U_q(sl_2)$, see Appendix A. Thus,
the so-obtained expression for the 
Hamiltonian $H$ is a simple function of $J^2$ and $J_3$. 

\section{A    deformed-boson $+$ $k$-fermion approach to fractional
supersymmetry}
\subsection{A deformed-boson $+$ $k$-fermion realization of $W_k$}
In this section, the main tools consist of $k$ pairs ($b(s)_-,b(s)_+$) with 
$s = 0, 1, \cdots, k-1$ of deformed bosons and a pair ($f_-,f_+$) 
of $k$-fermions. The operators $f_{\pm}$ satisfy (see Appendix B)
$$
 [ f_- , f_+ ]_q = 1, \quad
 f_-^k = 
 f_+^k = 0,
$$
and the operators $b(s)_{\pm}$ the commutation relation
$$
 [ b(s)_- , b(s)_+ ] = f_s(N),
 \eqno (23)
$$
where the functions $f_s$ with $s = 0, 1, \cdots, k-1$ and 
the operator $N$ occur in Eq.~(2). In addition, the pairs 
($f_-,f_+$) and ($b(s)_-,b(s)_+$) are two pairs of 
commuting operators and the operators $b(s)_{\pm}$ commute with the projection 
operators $\Pi_t$ with $s,t = 0, 1, \cdots, k-1$. We also introduce the 
linear combinations
$$
b_- := \sum_{s=0}^{k-1} b(s)_- \> \Pi_s, \quad 
b_+ := \sum_{s=0}^{k-1} b(s)_+ \> \Pi_s.
$$
It is immediate to verify that we have the commutation relation
$$
[ b_- , b_+ ] = \sum_{s=0}^{k-1} f_s(N) \> \Pi_s,
 \eqno (24)
$$
a companion of Eq.~(23).

   We are now in a situation to find a realization of the generators $X_-$,
$X_+$ and $K$ of the algebra $W_k$ in terms of the $b$'s and $f$'s. Let us
define 
the shift operators $X_-$ and $X_+$ by
$$
X_- := b_- \left( f_-   +    \frac{ f_+ ^{k-1}}{[[k - 1]]_q !} \right),
\eqno (25)
$$
$$
X_+ := b_+ \left( f_-   +    \frac{ f_+ ^{k-1}}{[[k - 1]]_q !} \right)^{k-1},
\eqno (26)
$$
where the new symbol $[[ \ ]]_q$ is defined by
$$
[[X]]_q := {1 - q^X \over 1 - q} 
$$
with $X$ an arbitrary operator or number
and where the $q$-deformed factorial is given by
$$
\lbrack\lbrack n \rbrack\rbrack_q ! := 
\lbrack\lbrack 1 \rbrack\rbrack_q \>
\lbrack\lbrack 2 \rbrack\rbrack_q \> \cdots \> 
\lbrack\lbrack n \rbrack\rbrack_q 
$$
for  $n \in {\bf N}^*$  (and   
$\lbrack\lbrack 0 \rbrack\rbrack_q ! := 1$). 
It is also always possible to find a representation 
for which the relation $X_-^{\dagger} = X_+$ holds (see Appendix C). 
Furthermore, we define the grading operator $K$ by
$$
K := [ f_- , f_+ ].
\eqno (27)
$$
In view of the remarkable property
$$
\left( f_- + \frac{f_+^{k-1}}{[[k-1]]_q!} \right)^k = 1,
$$ 
we obtain 
$$
[X_- , X_+] = 
[b_- , b_+].
\eqno (28)
$$
Equations (24) and (28) show that Eq.~(2a) is satisfied. 
It can be checked also 
that the operators $X_-$, $X_+$ and $K$ satisfy Eqs.~(2c) and (2e). Of course, 
Eqs.~(2b) and (2d) have to be considered as postulates. However, note that the 
operator $N$ is formally given in terms of the $b$'s by
$$
F(N+1) = b_- b_+ = \sum_{s=0}^{k-1} b(s)_- b(s)_+ \> \Pi_{s},
$$
$$
F(N)   = b_+ b_- = \sum_{s=0}^{k-1} b(s)_+ b(s)_- \> \Pi_{s},
$$
with the help of the structure function $F$ introduced in Sec.~2. We thus have
a 
realization of the generalized Weyl-Heisenberg algebra $W_k$ by multilinear
forms 
involving $k$ pairs ($b(s)_- , b(s)_+$) of deformed-boson 
operators ($s = 0, 1, \cdots, k-1$) and one pair ($f_- , f_+$) 
of $k$-fermion operators.

\subsection{The resulting Hamiltonian}
The supercharges $Q_-$ and $Q_+$ can be expressed by means of the 
deformed-bosons and k-fermions. By using the identity
$$
\Pi_s \left( f_-   +    \frac{ f_+ ^{k-1}}{[[k - 1]]_q !} \right)^{n} =
      \left( f_-   +    \frac{ f_+ ^{k-1}}{[[k - 1]]_q !} \right)^{n}
\Pi_{s+n},
$$
with $s = 0, 1, \cdots, k-1$ and $n \in {\bf N}$, 
Eqs.~(12) can be rewritten as
$$
Q_- = \left( f_-   +    \frac{ f_+ ^{k-1}}{[[k - 1]]_q !} \right) 
\> \sum_{s=1}^{k-1} b(s)_-   \> \Pi_{s+1},
$$
$$
Q_+ = \left( f_-   +    \frac{ f_+ ^{k-1}}{[[k - 1]]_q !} \right)^{k-1} 
\> \sum_{s=1}^{k-1} b(s+1)_+ \> \Pi_s,
$$
with the convention $b(k)_+ = b(0)_+$. Then, the supersymmetric 
Hamiltonian $H$ given by Eq.~(17) assumes the form
$$
 H = (k-1)
       \sum_{s =   0}^{k-1} F_s(N)                               \Pi_{s}   
   - \sum_{s = k-2}^{k  } 
       \sum_{t =   2}^{s-1} (t - 1) \> f_{t}(N - s + t )         \Pi_{s}
$$
$$
   - \sum_{s =   1}^{k-1} 
       \sum_{t =   s}^{k-1} (t - k) \> f_{t}(N - s + t )         \Pi_{s},
$$ 
in terms of the operators $b(s)_{\pm}$, 
            the projection operators
$\Pi_s$ (that may be written with $k$-fermion operators),
            the structure functions $F_s$  
        and the structure constants $f_s$ with $s = 0, 1, \cdots, k-1$.

\section{The fractional supersymmetric oscillator}
\subsection{A special case of $W_k$}
In this section, we deal with the particular case where $f_s = 1$ and the
deformed 
bosons $b(s)_{\pm} \equiv b_{\pm}$ are independent of $s$ 
with $s = 0, 1, \cdots, k-1$. We thus end 
up with a pair ($b_- , b_+$) of ordinary bosons, satisfying $[b_-,b_+] = 1$, 
    and a pair ($f_- , f_+$) of 
$k$-fermions. The ordinary bosons $b_{\pm}$ and the $k$-fermions $f_{\pm}$ may
be considered as 
originating from the decomposition of a pair of $Q$-uons when $Q$ goes to the 
root of unity $q$ (see Appendix B).

Here,the two operators $X_-$ and $X_+$ are given by Eqs.~(25) and (26), where
now
$b_{\pm}$ are ordinary boson operators. They satisfy the commutation relation
$[X_- , X_+] = 1$. Then, the number operator $N$ may defined by
$$
N := X_+ X_-, 
\eqno (29{\rm a})
$$
which is amenable to the form
$$
N  = b_+ b_-.
\eqno (29{\rm b})
$$
Finally, the  grading  operator  $K$  is still defined by Eq.~(27).
We can check that the operators $X_-$, $X_+$, $N$ and $K$ 
so-defined generate the generalized
Weyl-Heisenberg algebra $W_k$ defined by  Eq.~(2)  with $f_s = 1$ for
$s = 0, 1, \cdots, k-1$. The latter algebra $W_k$ can thus be realized 
with multilinear forms involving ordinary boson operators 
$b_{\pm}$ and $k$-fermion operators $f_{\pm}$. 

\subsection{The resulting fractional supersymmetric oscillator}
\noindent The supercharge operators $Q_-$ and $Q_+$ as well as the Hamiltonian
$H$ 
associated to the algebra $W_k$ introduced in Sec.~4.2
(in terms of the operators $b_-$, $b_+$, $f_-$ and $f_+$)
can be constructed according to the prescriptions given in Sec.~3. This
leads to the expression
$$
H = (k-1) b_+ b_-  +  (k-1) \sum_{s=0}^{k-1} ( s + 1 - \frac{1}{2} k )
\Pi_{k-s}
$$
to be compared with Eq.~(21). 

Most of the 
properties of the Hamiltonian $H$ are essentially
the same as the ones given above for the Hamiltonian (21). 
In particular, we can write
$$
H = \sum_{m = 1}^{k} H_m \Pi_m, \quad 
                     H_m := (k-1) \left( b_+ b_- + \frac{1}{2} k + 1 - m
\right)
$$
and  thus $H$  is  a linear combination of
projection operators with coefficients $H_m$ corresponding to isospectral
Hamiltonians
(remember that $\Pi_k := \Pi_0$). 

To close this section, 
let us mention that the fractional supercoherent state 
$| z , \theta )$ defined in Appendix B is a coherent state 
corresponding to the Hamiltonian $H$. As a point of fact, we  
can check that the action of the evolution operator
$\exp (- {\rm i} H t)$ on the state $| z , \theta )$ gives
$$
\exp (- {\rm i} H t) \> | z , \theta ) =
\exp \left[ - \frac{{\rm i}}{2} (k-1) (k+2) t \right] 
                     \> | {\rm e}^{-{\rm i} (k-1) t} z , 
                          {\rm e}^{+{\rm i} (k-1) t} \theta ),
$$
i.e., another fractional supercoherent state. 

\subsection{Examples}
\noindent {Example 1.} As a first example, we take $k=2$, i.e., $q=-1$. Then,
the
operators
$$
X_{\pm} := b_{\pm} \left( f_-   +    f_+ \right) 
$$
\noindent and the operators $K$ and $N$, see Eqs.~(27) and (29), are defined
in terms of bilinear forms of ordinary bosons 
$(b_- , b_+)$ and             ordinary fermions
$(f_- , f_+)$. The operators $X_-$, $X_+$, $N$ and $K$ satisfy 
$$
 [X_- , X_+     ]   = 1, \quad 
 [N   , X_{\pm} ]   = {\pm} X_{\pm}, 
$$
$$
 [K   , X_{\pm} ]_+ = 0, \quad
 [K   , N       ]   = 0, \quad K^2 = 1,
$$
\noindent which reflect bosonic  and  fermionic 
degrees of freedom, the bosonic degree corresponding to the triplet 
($X_- ,  X_+ , N$) and the fermionic degree to the Klein involution 
operator $K$. The projection operators
$$
  \Pi_0 = 1 - f_+ f_-, \quad
  \Pi_1 =     f_+ f_-
$$
\noindent are here simple chirality operators and the supercharges
$$
  Q_- = b_- f_+, \quad
  Q_+ = b_+ f_-
$$
\noindent have the property
$$
Q_- ^2 = 
Q_+ ^2 = 0. 
$$
\noindent The Hamiltonian $H$ assumes the form
$$
H = [ Q_- , Q_+ ]_+ 
$$
\noindent which can be rewritten as
$$
H = b_+ b_- \Pi_0   +   b_- b_+ \Pi_1. 
$$
\noindent It is clear that the self-adjoint operator $H$ commutes with $ Q_- $ 
and   $Q_+$ and we recover that 
$Q_-$ , $Q_+$ and $H$ span the Lie superalgebra 
$s \ell (1/1)$. We have
$$
H = b_+ b_-    +   f_+ f_-
$$
\noindent so that $H$ acts on the $Z_2$-graded space 
${\cal F} = {\cal F}_0 \oplus {\cal F}_1$. 
The operator $H$ corresponds to the ordinary or $Z_2$-graded 
supersymmetric oscillator whose energy spectrum $E$ is (in a symbolic way)
$$
E = 1 \oplus 2 \oplus 2 \oplus \cdots
$$ 
\noindent with equally spaced levels, the ground state  being  a 
singlet (denoted by 1) and all the excited states (viz., an infinite sequence) 
being doublets
(denoted by 2). Finally, note that the fractional supercoherent state
$ | z , \theta ) $ of Appendix B with $k=2$ is a coherent state for the
Hamiltonian $H$ (see also Ref.~[46]).

\noindent {Example 2.} We continue with $k=3$, i.e., 
$$
q = \exp \left( \frac{2 \pi {\rm i}}{3} \right).
$$
In this case, we have 
$$
X_- = b_- \left( f_- - q f_+^2 \right),
$$
$$
X_+ = b_+ \left( f_+ + f_-^2 + q^2 f_+^2 f_- \right),
$$
\noindent and $K$ and $N$ as given by (27) and (29), 
where here $( b_- , b_+ )$ are ordinary bosons 
and $( f_- , f_+ )$ are  $3$-fermions. We hence have 
$$
 [ X_- , X_+ ] = 1, \quad
 [ N , X_{\pm} ] = {\pm} X_{\pm},
$$
$$
 [ K , X_+ ]_q  = [ K , X_- ]_{\bar q} = 0, \quad
 [K , N] = 0,         \quad  K^3 = 1.
$$
\noindent Our general definitions can be specialized to
\begin{eqnarray*}
\Pi_0 &=& 1 + (q-1) f_+ f_-  -  q    f_+ f_- f_+ f_- \\
\Pi_1 &=&       - q f_+ f_-  + (1+q) f_+ f_- f_+ f_- \\
\Pi_2 &=&           f_+ f_-  -       f_+ f_- f_+ f_-
\end{eqnarray*}
\noindent for the projection operators and to 
$$
Q_- = b_- f_+ \left( f_-^2 -  q f_+   \right) 
$$
$$ 
Q_+ = b_+     \left( f_-   -  q f_+^2 \right) f_- 
$$ 
\noindent for the supercharges with the property
$$
Q_- ^3 = 
Q_+ ^3 = 0. 
$$
\noindent By introducing the Hamiltonian $H$ via
$$
Q_- ^{2} Q_+     +   Q_- Q_+ Q_- 
                               +   Q_+ Q_- ^{2}
                               =   Q_- H
$$
\noindent we obtain
$$
H = \left( 2 b_+ b_-  -  1 \right) \Pi_0 +
    \left( 2 b_+ b_-  +  3 \right) \Pi_1 +
    \left( 2 b_+ b_-  +  1 \right) \Pi_2
$$
\noindent which acts on the $Z_3$-graded space 
${\cal F} = {\cal F}_0 \oplus {\cal F}_1 \oplus {\cal F}_2$ 
and can be rewritten as 
$$
H = 2 b_+ b_-  -  1  +  2(1 - 2q) f_+ f_-  +  2(1 + 2 q) f_+ f_- f_+ f_-
$$
\noindent in terms of boson and 3-fermion operators.
We can check that the self-adjoint operator $H$ commutes with $ Q_- $ 
and   $Q_+$. The energy spectrum of $H$ reads 
$$
E = 1 \oplus 2 \oplus 3 \oplus 3 \oplus \cdots.
$$ 
\noindent It contains equally spaced levels with a 
nondegenerate ground state (denoted as 1), a doubly 
degenerate first excited state (denoted as 2) and an infinite
sequence of triply degenerate excited states (denoted as 3). 

\section{Differential realizations}
In this section, we consider the case of the algebra $W_k$ defined by
Eqs.~(2b)-(2e) and Eq.~(8) with $c_0 = 1$ and $c_s = c \delta (s,1)$, 
$c \in {\bf R}$, for $s = 1, 2, \cdots, k-1$. In other words, we have
$$
 [X_- , X_+] = 1 + c K, \quad K^k = 1,
\eqno (30{\rm a})
$$
$$
 [K , X_+]_q = [K , X_-]_{\bar q} = 0,
\eqno (30{\rm b}) 
$$
which corresponds to the $C_{\lambda}$-extended oscillator. The operators
$X_-$, $X_+$ and $K$ can be realized in terms of a bosonic variable $x$ and
its derivative $ \frac{d}{dx} $ satisfying
$$
[ \frac{d}{dx}  , x] = 1
$$
and a $k$-fermionic variable (or generalized 
Grassmann variable) $\theta$ and its
derivative $ \frac{d}{d \theta} $ satisfying$^{21,32}$ 
(see also Refs.~[22-28])
$$
[ \frac{d}{d \theta}  , \theta]_{\bar q} = 1, \quad 
{\theta}^k = \left( { \frac{d}{d \theta} } \right)^k=0.
$$
Of course, the sets $\{ x ,  \frac{d}{dx}  \}$ and $\{ \theta ,  \frac{d}{d
\theta}  \}$
commute. It is a simple matter of calculation to derive the two following
identities
$$
\left(  \frac{d}{d \theta}  + \frac{\theta^{k-1}}{[[k-1]]_{\bar q}!} \right)^k
= 1
$$
and
$$
(  \frac{d}{d \theta}  \theta - \theta  \frac{d}{d \theta}  )^k = 1,
$$
which are essential for the realizations given below. 

As a first realization, we can show that the shift operators
$$
X_- =  \frac{d}{dx}  \left(  \frac{d}{d \theta}  +
\frac{\theta^{k-1}}{[[k-1]]_{\bar q}!} \right)^{k-1} 
    - \frac{c}{x} \theta,
$$
$$
X_+ = x \left(  \frac{d}{d \theta}  + \frac{\theta^{k-1}}{[[k-1]]_{\bar q}!}
\right),
$$
and the Witten grading operator
$$
K = [  \frac{d}{d \theta}  , \theta ] 
$$
satisfy Eqs.~(30). This realization of $X_-$, $X_+$ and $K$ 
clearly exibits the bosonic and $k$-fermionic degrees of 
liberty via the sets $\{ x ,  \frac{d}{dx}  \}$ and $\{ \theta ,  \frac{d}{d
\theta}  \}$,
respectively. In the particular case $k=2$, the $k$-fermionic 
variable $\theta$ is an ordinary Grassmann variable and the 
supercharge operators $Q_-$ and $Q_+$ take the simple form
$$
Q_- = \left(  \frac{d}{dx}  - \frac{c}{x} \right) \theta 
\eqno (31)
$$
$$
Q_+ = x  \frac{d}{d \theta}.
\eqno (32)
$$
(Note that the latter realisation for $Q_-$ and $Q_+$ is
valid for $k = 3$ too.)

Another possible realization of $X_-$ and $X_+$ 
for arbitrary $k$ is 
$$
X_- = P \left(  \frac{d}{d \theta}  + \frac{\theta^{k-1}}{[[k-1]]_{\bar q}!}
\right)^{k-1} 
    - \frac{c}{x} \theta,
$$
$$
X_+ = X \left(  \frac{d}{d \theta}  + \frac{\theta^{k-1}}{[[k-1]]_{\bar q}!}
\right),
$$
where $P$ and $X$ are the two canonically conjugated quantities
$$
P := \frac{1}{\sqrt{2}} \left(  x + \frac{d}{dx} - \frac{c}{2x} K \right)
$$
and
$$
X := \frac{1}{\sqrt{2}} \left(  x - \frac{d}{dx} + \frac{c}{2x} K \right).
$$
This realization is more convenient for a Schr\"odinger type approach
to the supersymmetric Hamiltonian 
$H$. According to Eq.~(17), 
we can derive an Hamiltonian 
$H$ involving bosonic and $k$-fermionic degrees of freedom.
To illustrate this point, let us 
continue with the particular case $k=2$. It can be seen that 
the supercharge operators (31) 
and (32) must be replaced by 
$$
Q_- = \left( P - \frac{c}{X} \right) \theta
$$
$$
Q_+ = X  \frac{d}{d \theta}.
$$
(Note the formal character of $Q_-$ since the 
definition of $Q_-$ lies on the existence 
of an inverse for the operator $X$.) Then, 
we obtain the following Hamiltonian 
$$
H = - \frac{1}{2} \left[ \left( \frac{d}{dx} - \frac{c}{2 x} K \right)^2 
    - x^2 + K + c (1 + K) \right].
$$
For $c=0$, we have (cf.~Ref.~[6])
$$
H = - \frac{1}{2} \frac{d^2}{dx^2} 
    + \frac{1}{2} x^2  
    - \frac{1}{2} K 
$$
that is the Hamiltonian for an ordinary super-oscillator, i.e., a 
$Z_2$-graded supersymmetric oscillator. Here, the bosonic character
arises from the bosonic variable $x$ and the fermionic    character
       from the ordinary Grassmann variable $\theta$ in $K$. 

\section{Concluding remarks}
A first facet of this work concerns an approach of 
${\cal N}=2$ FSSQM of order $k$ 
($k \in {\bf N} \setminus \{ 0 , 1 \}$)
in $D = 1+1$ dimensions
through a generalized Weyl-Heisenberg algebra $W_k$ which is an extension of 
the Calogero-Vasiliev$^{43}$ algebra. We 
have shown how the algebra $W_k$ is
connected to the quantum algebra $U_q(sl_2)$ with $q^k = 1$. This 
approach  of  FSSQM, in the spirite of the pioneer works in 
Refs.~[30,31], 
differs from the one developed in 
Refs.~[21-27] via the introduction of two
degrees of freedom, a bosonic one and a para-fermionic one. At first glance,
our approach seems to be of an entirely bosonic character. However, the
para-fermionic or $k$-fermionic character is hidden behind the (Klein-Witten)
operator $K$. This operator ensures a $Z_k$-grading of the Hilbert space 
${\cal F}$ of the physical states according to the decomposition 
${\cal F} = \bigoplus_{s=0}^{k-1} {\cal F}_s$. The generators of $W_k$ 
(and consequently of $U_q(sl_2)$) have been used for constructing a general
fractional supersymmetric Hamiltonian $H$ which is a linear combination of
projection operators on the subspaces ${\cal F}_s$ ($s = 0, 1, \cdots, k-1$),
the coefficients of which being isospectral Hamiltonians. The general
Hamiltonian $H$ covers the particular case of the fractional supersymmetric 
oscillator.

A second facet of this paper is devoted to a $Q$-uon approach of 
${\cal N} = 2$ FSSQM of 
order $k$ with $Q$ going to $q = \exp (2 \pi {\rm i} / k)$.  The bosonic and
$k$-fermionic degrees of freedom are present since
the very beginning, a situation which parallels the {\em \`a la} Rubakov
and Spiridonov$^{21,22}$
construction of para-supersymmetric quantum mechanics.   
Indeed, the  
$Q$-{\em uon} $\to$ {\em boson} $+$ $k$-{\em fermion decomposition} obtained
when $Q \leadsto q$ has been exploited for building a realization of $W_k$ 
corresponding to the fractional supersymmetric oscillator. This approach
of FSSQM is especially appropriate for deriving the  fractional
supercoherent states associated to this fractional
supersymmetric oscillator. In addition, it is appropriate to the writing of 
supercharges and fractional supersymmetric Hamiltonians in terms of
ordinary bosonic variables and generalized Grassmann variables, as shown
with the specific differential realizations of Sec.~6.

The two approaches of FSSQM developed in this paper are obviously
complementary. In this direction, it is to be emphasized 
that this work might be useful for generating 
isospectral Hamiltonians  for  exactly integrable potentials 
and for constructing their coherent states.

Finally, two comments of a group-theoretical nature are in order. First, we
have
shown here
that supercharges and fractional supersymmetric Hamiltonians can 
be expressed from the generators of $U_q(sl_2)$, with $q$ a $k$-th root of
unity, in a way independent of the 
representations (i), (ii) and (iii) of Appendix A chosen for the quantum
algebra $U_q(sl_2)$. This approach is different from the one in 
Refs.~[32,35]   
where nilpotent and cyclic representations  of  $U_q(sl_2)$, with 
$q^2$ being a root of unity, are separately considered for an investigation 
of ${\cal N} = 2$ FSSQM  in $D= 1+1$ dimensions. Second, the algebra
$U_q(sl_2)$
has not to be confused with the algebra spanned by the supercharges $Q_-$ and
$Q_+$
and the Hamiltonian $H$. The latter algebra coincides with 
the $Z_2$-graded Lie algebra $sl(1/1)$ for $q=-1$, i.e., 
$k=2$, in the case of ${\cal N}=2$ SSQM. An 
open question is to find the algebra spanned by $Q_-$, $Q_+$
and $H$ for $k \ge 3$ in the case of 
${\cal N}=2$ FSSQM. In an other terminology, can ${\cal N}=2$ FSSQM
of order $k$ 
be described by a $q$-deformed algebra (with $q^k = 1$) that
gives back $sl(1/1)$ for $q=-1$~?
It is hoped that the results in this paper shall shed light
on this question. 

\section*{Acknowledgments}

One of the authors (M.~D.) would like to thank 
the Institut de Physique Nucl\'eaire de Lyon 
for the kind hospitality extended to him 
at various stages (during 1999-2001) 
of the development of this work.

\section*{Appendix A: Connection between $W_k$ and $U_q(sl_2)$}
Let us now show that the quantum algebra $U_q(sl_2)$, 
with $q$ being the $k$-th root of unity
given by (1), turns out to be a particular form of $W_k$. The algebra
$U_q(sl_2)$ is spanned by the generators
$J_-$,  $J_+$,  $q^{J_3}$  and  $q^{-J_3}$
that satisfy the relationships 
$$
[J_+ , J_-] =  [2 J_3]_q,
$$ 
$$
q^{J_3} J_+ q^{-J_3} =       q  J_+, \quad
q^{J_3} J_- q^{-J_3} = {\bar q} J_-,
$$
$$
q^{J_3} q^{-J_3} = q^{-J_3} q^{J_3} = 1.  
$$
It is straightforward to prove that the operator
$$
J^2 := J_- J_+  +  \frac{q^{+1} q^{2J_3} + q^{-1} q^{-2J_3}} {(q - q^{-1})^2}
$$ 
or
$$
J^2 := J_+ J_-  +  \frac{q^{-1} q^{2J_3} + q^{+1} q^{-2J_3}} {(q - q^{-1})^2}
$$ 
is an invariant of $U_q(sl_2)$. In view of Eq.~(1), the operators $J_-^k$,
$J_+^k$, 
$( q^{ J_3} )^k$, and 
$( q^{-J_3} )^k$ belong, likewise $J^2$, to the center of $U_q(sl_2)$. 

In the case where the deformation parameter $q$ is 
a root of unity,
the representation theory of $U_q(sl_2)$ is richer than
the one for $q$ generic. The algebra $U_q(sl_2)$ admits 
finite-dimensional 
representations of dimension $k$ such that
$$
J_-^k = A, \quad 
J_+^k = B,
$$
where $A$ and $B$ are constant matrices. Three types of representations
have been studied in the literature:$^{45}$

(i)   $A = B = 0$ (nilpotent representations),

(ii)  $A = B = 1$ (cyclic or periodic representations),

(iii) $A = 0$ and $B = 1$ or 
      $A = 1$ and $B = 0$
      (semi-periodic representations).
      
\noindent Indeed, the realization of FSSQM
based on $U_q(sl_2)$  does not depend of the choice (i), (ii) or (iii) 
in contrast with the work in 
Ref.~[32] where nilpotent representations 
corresponding to the choice (i) 
were considered. The only
important ingredient is to take 
$$
\left( q^{J_3} \right)^k = 1
$$
that ensures a $Z_k$-grading of the Hilbertean 
representation space of $U_q(sl_2)$. 

The contact with the algebra $W_k$ is established by putting
$$
X_{\pm} := J_{\pm}, \quad
N   := J_3,         \quad
K   := q^{J_3},
$$
and by using the definition (3) of $\Pi_s$ as function of $K$. 
Here, the operator $\Pi_s$ is a projection operator on the subspace,
of the representation space of $U_q(sl_2)$,
corresponding to a given eigenvalue of $J_3$. 
It is easy to check that the
operators $X_-$, $X_+$, $N$ and $K$ satisfy Eqs.~(2) with
$$
f_s(N) = - [2s]_q = - \frac{\sin \frac{4 \pi s}{k}}{\sin \frac{2 \pi}{k}}
$$
for $s = 0, 1, \cdots, k-1$. The quantum algebra 
$U_q(sl_2)$, with $q$ given by (1), then appears as a further particular 
case of the generalized Weyl-Heisenberg algebra 
$W_k$.

\section*{Appendix B: The $Q$-uon $\to$ boson $+$ $k$-fermion decomposition}
We shall limit ourselves to give an outline of this decomposition (see 
      Dunne {\em et al.}$^{47}$
and Mansour {\em et al.}$^{48}$ for an alternative and more
rigorous mathematical presentation based on the isomorphism between the
braided $Z$-line and the $(z , \theta)$-superspace). We start from a 
$Q$-uon algebra spanned by three operators $a_-$, $a_+$ and $N_a$ 
satisfying the relationships$^{16}$
(see also Refs.~[17-20])
$$
[a_- , a_+]_Q = 1, \quad
[ N_a , a_{\pm} ] = {\pm} a_{\pm},
$$
where $Q$ is generic (a real number different from zero). The
action of the operators $a_-$, $a_+$ and $N_a$ on a Fock space
${\cal F} := \{ | n \rangle : n \in {\bf N} \}$ is given by 
$$
N_a | n \rangle = n | n \rangle,
$$
and
$$
a_- | n \rangle = ([[n + \sigma - \frac{1}{2}]]_Q)^{\alpha} \> | n - 1 \rangle,
$$
$$
a_+ | n \rangle = ([[n + \sigma + \frac{1}{2}]]_Q)^{\beta}  \> | n + 1 \rangle,
$$ 
where $\sigma = \frac{1}{2}$ and 
$\alpha + \beta = 1$ with $0 \leq \alpha \leq 1$ and 
$0 \leq \beta \leq 1$. For $\alpha = \beta = \frac{1}{2}$, 
let us consider the $Q$-deformed Glauber coherent 
state$^{16}$ (see also Ref.~[49])
$$
| Z ) := \sum_{n=0}^{\infty} 
\frac{ \left( Z a_+ \right)^n }{ [[n]]_Q! } \> | 0 \rangle
       = \sum_{n=0}^{\infty} 
   \frac{ Z^n }{ ([[n]]_Q!)^{\frac{1}{2}} } \> | n \rangle
$$ 
\noindent (with $Z \in {\bf C}$). If we do the replacement 
$$
Q \leadsto q := \exp \left( \frac{2 \pi {\rm i}}{k} \right), \quad 
k \in {\bf N} \setminus \{ 0,1 \},
$$
we have 
$\lbrack\lbrack k \rbrack\rbrack_Q ! \to 
 \lbrack\lbrack k \rbrack\rbrack_q ! = 0$. Therefore,
in order to give a sense to $| Z )$ for $Q \leadsto q$, we have to do the
replacement
$$
a_+ \leadsto f_+ \quad {\rm with} \quad f_+^k = 0
$$
\noindent and, for the sake of symmetry, a similar 
replacement for $a_-$. We 
thus end up with what we call a $k$-fermionic algebra
spanned by the operators  $f_-$, $f_+$ and $N_f \equiv N_a$ completed by the
adjoints
$f_+^{\dagger}$ and 
$f_-^{\dagger}$ of $f_+$ and $f_-$, 
respectively.$^{28,29}$ The defining
relations for the $k$-fermionic algebra are
$$
 [ f_- , f_+ ]_q = 1, \quad
 [ N_f , f_{\pm} ] = {\pm} f_{\pm}, \quad
 f_-^k = 
 f_+^k = 0,
$$
and similar relations for $f_+^{\dagger}$ and 
                          $f_-^{\dagger}$. The case $k=2$ corresponds
to ordinary fermion operators and the case $k \to \infty$ 
to ordinary boson   operators. The $k$-fermions are objects interpolating
between
fermions and bosons. They share some properties with the 
para-fermions$^{21,22,24}$
and the anyons as introduced by 
Goldin {\em et al.} [8]
(see also Ref.~[7]). If we define 
$$
b_{\pm} := \lim_{Q \leadsto q} \frac{    a_{\pm}^k    }
                               { \left( [[k]]_Q ! \right)^{\frac{1}{2}}} 
$$
we obtain 
 $$
 [ b_- , b_+ ] = 1
 $$
so that the operators $b_-$ and $b_+$ can be considered as ordinary boson
operators. 
This is at the root of the two following results.$^{28}$

As a first result, the set $\{ a_- , a_+ \}$ gives rise, for $Q \leadsto q$, 
to two commuting sets: 
The set $\{ b_- , b_+ \}$ of boson operators and the set of $k$-fermion
operators $\{ f_- , f_+ \}$. As a second result, this decomposition leads 
to the replacement of the $Q$-deformed coherent state $ | Z ) $ by the 
so-called fractional supercoherent state
$$
| z , {\theta} ) := \sum_{n=0}^{\infty} 
                    \sum_{s=0}^{k-1} 
                    \frac{ \theta^s }{ ([[s]]_q!)^{\frac{1}{2}} }
                    \> \frac{ z^n }{ \sqrt{n!} }
                    \> | kn + s \rangle,
$$
\noindent where $z$ is a (bosonic) complex variable and $\theta$ a 
($k$-fermionic) generalized Grassmann 
variable$^{21,24,32,50}$
with $\theta^k = 0$. The fractional 
supercoherent state $| z , {\theta} )$ 
is an eigenvector of the product $f_- b_-$ with the eigenvalue 
$z \theta$. The state $| z^k , {\theta} )$ can be seen to 
be a linear combination of the coherent states introduced by 
Vourdas$^{51}$
with coefficients in the generalized Grassmann algebra spanned by $\theta$ and
the derivative $ \frac{d}{d \theta} $. 
   
In the case $k=2$, the fractional supercoherent state $| z , {\theta} )$
turns out to be a coherent state for the ordinary (or $Z_2$-graded) 
supersymmetric oscillator.$^{46}$ For $k \ge 3$, the state $| z , {\theta} )$
is a coherent state for the $Z_k$-graded supersymmetric oscillator (see
Sec.~5). 

\section*{Appendix C: Actions on the space ${\cal F}$}
Equation (23) is satisfied by
$$
b(s)_- b(s)_+ = F_s(N + 1), \quad
b(s)_+ b(s)_- = F_s(N),
$$
where the structure functions $F_s$ are connected 
to the structure constants $f_s$ via
$$
F_s(N + 1) - F_s(N) = f_s(N) 
$$
and to the the structure function $F$ via
$$
F(N) = \sum_{s=0}^{k-1} F_s(N) \> \Pi_s
$$
(see Sec.~2 for the definition of $f_s$ and $F$).

Let us consider the operators $X_-$ and $X_+$ defined by Eqs.~(25) 
and (26) 
and acting on the Hilbert-Fock space ${\cal F}$ (see Eqs.~(5)). We choose the
action 
of the constituent operators 
$b_{\pm}$ and $f_{\pm}$ on the state $| k n + s \rangle$
to be given by
$$
b_-    | kn + s \rangle = 
b(s)_- | kn + s \rangle = 
\sqrt {F_s(n + \sigma - \frac{1}{2})} \> 
| k(n - 1)+s \rangle,
$$
$$
b_+    | kn + s \rangle = 
b(s)_+ | kn + s \rangle = 
\sqrt {F_s(n + \sigma + \frac{1}{2})} \> 
| k(n + 1)+s \rangle,
$$   
and
$$
f_- | kn + s \rangle = | kn + s-1 \rangle, \quad
f_- | kn \rangle = 0,
$$
$$
f_+ | kn + s \rangle = [[s + 1]]_q \> | kn + s+1 \rangle, \quad
f_+ | kn + k - 1 \rangle = 0,
$$
where $n \in {\bf N}$ and $s= 0, 1, \cdots, k-1$. The action of 
$b_{\pm}$ is standard and the action of $f_{\pm}$ corresponds 
to $\alpha = 0$ and $\beta = 1$ (see Appendix B).
Then, we can show that the relationships (6) are satisfied. In 
this representation, it is easy to prove that the
Hermitean conjugation relation $X_-^{\dagger} = X_+$ 
is true.

  \end{document}